\documentclass[letterpaper,twocolumn]{article}
\usepackage[raggedright]{titlesec}
\usepackage{amsmath,graphicx}
\usepackage{balance}

\graphicspath{{../imgs/}}

\date{}

\titlespacing{\section}{0pt}{*2.0}{*2.0}
\titlespacing{\subsection}{0pt}{*1.8}{*1.8}

\begin{document}

% Change to your title
\title{\Large\textbf{Prediction of Future Hospital Admissions -- What is the Tradeoff Between Specificity and Accuracy?}\normalsize}

%Chane authors, institutions and emails
\author {Ieva Vasiljeva and Ognjen Arandjelovi\'c$^\dag$\\~\\School of Computer Science\\University of St Andrews\\North Hough\\St Andrews KY16 9SX\\Scotland, Fife\\United Kingdom\\$^\dag$\texttt{ognjen.arandjelovic@gmail.com}}
\maketitle 

\thispagestyle{empty}

\begin{center}
\large\textbf{Abstract}
\end{center}

\vspace{2mm}
Large amounts of electronic medical records collected by hospitals across the developed world offer unprecedented possibilities for knowledge discovery using computer based data mining and machine learning. Notwithstanding significant research efforts, the use of this data in the prediction of disease development has largely been disappointing. In this paper we examine in detail a recently proposed method which has in preliminary experiments demonstrated highly promising results on real-world data. We scrutinize the authors' claims that the proposed model is scalable and investigate whether the tradeoff between prediction specificity (i.e.\ the ability of the model to predict a wide number of different ailments) and accuracy (i.e.\ the ability of the model to make the correct prediction) is practically viable. Our experiments conducted on a data corpus of nearly 3,000,000 admissions support the authors' expectations and demonstrate that the high prediction accuracy is maintained well even when the number of admission types explicitly included in the model is increased to account for 98\% of all admissions in the corpus. Thus several promising directions for future work are highlighted.

\medskip
\noindent

\section{Introduction}\label{s:intro}
The concept of prognosis is of pervasive importance in medicine. As the very etymology of the word suggests (Greek \textit{progn\={o}sis}, \textit{pro}-`before' + \textit{gign\={o}skein} `know'), prognosis concerns the \emph{prediction} of medical outcomes. Examples include disease progression patterns, different aspects of life quality, mortality chances, and many others. Already recognized in the antiquity (amongst others by Ancient Egyptians, Sumerians, and Greeks) the paradigm of prognosis -- that is to say the underlying methodology -- has changed dramatically. Initially predicated on highly limited and bias prone experience of practising physicians, the last couple of centuries have witnessed the development of a rigorous framework for collecting, interpreting, and using evidence -- this is now widely referred to as evidence based medicine (or better yet, science based medicine~\cite{GorsNove2014}).

%Considering the aforementioned trend it is unsurprising\ that
Recent advances in computing are promising to effect a major change in prognostic practice.  In particular, the ability to record, store, and send across large distances massive amounts of patient data, together with major breakthroughs in machine learning and data mining, open the possibility of using artificial intelligence to analyse corpora far larger than ever before. The particular focus of the present paper is on the use of electronic medical records for the prediction of future complications likely to be experienced by a specific patient. Such predictions can aid in the understanding of disease aetiology, patient incentivization, and the allocation of hospital resources~\cite{Aran2015g}.

Considering the impact that accurate prediction of this type would have on finances, population health at large, and individual patients' well-being, it is unsurprising that the challenge of developing suitable models has already attracted significant research efforts. Broadly categorized, these fall under two umbrellas: (i) methods which involve explicit physiological modelling relevant to a specific condition~\cite{ToppPromdeVrMiur+2000,DeGaHardBeckAbuR+2008,YeIsamBarh2012}, and (ii) methods which adopt a holistic strategy and approach the problem from a data-driven perspective, modelling the overall state of a patient's health~\cite{BottAyliMaje2006,KansEnglSalaKage+2011,Aran2015g}. An obvious limitation of the former group of methods is that by their very nature they are not readily generalizable. Moreover, it is likely that a restricted view of a single condition in isolation is inadequate to model many modern diseases which have numerous comorbidities affecting a broad range of bodily systems~\cite{LongDago2011}. To give but one example, some of the more common comorbidities of diabetes mellitus include macro-vascular ailments such as coronary artery disease, myocardial infarction, stroke, congestive heart failure, and peripheral vascular disease, micro-vascular complications such as retinopathy, nephropathy, and neuropathy, metabolic disorders such as dyslipidemia, non-alcoholic fatty liver disease, and obesity etc. Thus, holistic modelling approaches appear more attractive in principle. However, the task of developing models which are flexible enough to provide practically sufficient specificity, yet constrained enough to be learnable from real-world data, is a major challenge. Indeed, until recently the performance of different methods described in the literature has been disappointing. The present paper is motivated by a recently proposed method which has demonstrated impressive and at first sight highly promising results in preliminary experiments~\cite{Aran2015g}.

The key ideas underlying the adopted method and a sketch of its main technical aspects are described in the next section. Our main aim here was to scrutinize the original authors' expectation that the method would scale well i.e.\ that the predictive performance of the method, reported with explicit modelling of the 30 most frequent admission types only, could be maintained as a greater number of admission types is included in the model as most practical applications would demand. The original paper did not investigate this; rather, the number of salient, explicitly modelled admission types was set in an \emph{ad hoc} manner to 30, explaining approximately 75\% of the data corpus~\cite{Aran2015g} (also see \cite{Aran2016}). If our expectation of performance deterioration with an increased number of explicitly modelled admission types is correct, and if the rate of deterioration is high, the model could end up being of little practical significance: on the one end of the parameter spectrum the model would provide high accuracy but insufficient specificity for its predictions to be practically useful, and on the other high specificity but poor accuracy for its predictions to be relied upon. Thus the present analysis is necessary before any practical use can be considered.

%Former not really generalizable.  Moreover limited in modelling comorbidities and most modern diseases have lots of these
%The latter are in principle more attractive.  However it is a difficult task.  recent review found the results in the literature to be poor.  Highly promising performance recently reported in ~\cite{Aran2015g}.  Method will be described in more detail in the next section but summary
%Authors claim that if more data is available more admission codes can be modelled explicitly or greater granularity
%However they did not investigate this and in particular how the method behaves as the number of salient codes is changed.  Clearly there is a tradeoff

\section{Methods}\label{s:meth}
In the present paper we conduct our investigation using the method recently proposed in~\cite{Aran2015c,Aran2015g}.  Our choice was motivated by its high predictive accuracy demonstrated on a large real-world data corpus. For the sake of completeness the key ideas and the main technical elements underlying the adopted method are summarized next; for comprehensive detail and a related discussion the reader is referred to the original publication.

\subsection{Adopted model overview}\label{ss:model}
\begin{figure}
  \vspace{-0pt}
  \centering
  \includegraphics[width=0.99\columnwidth]{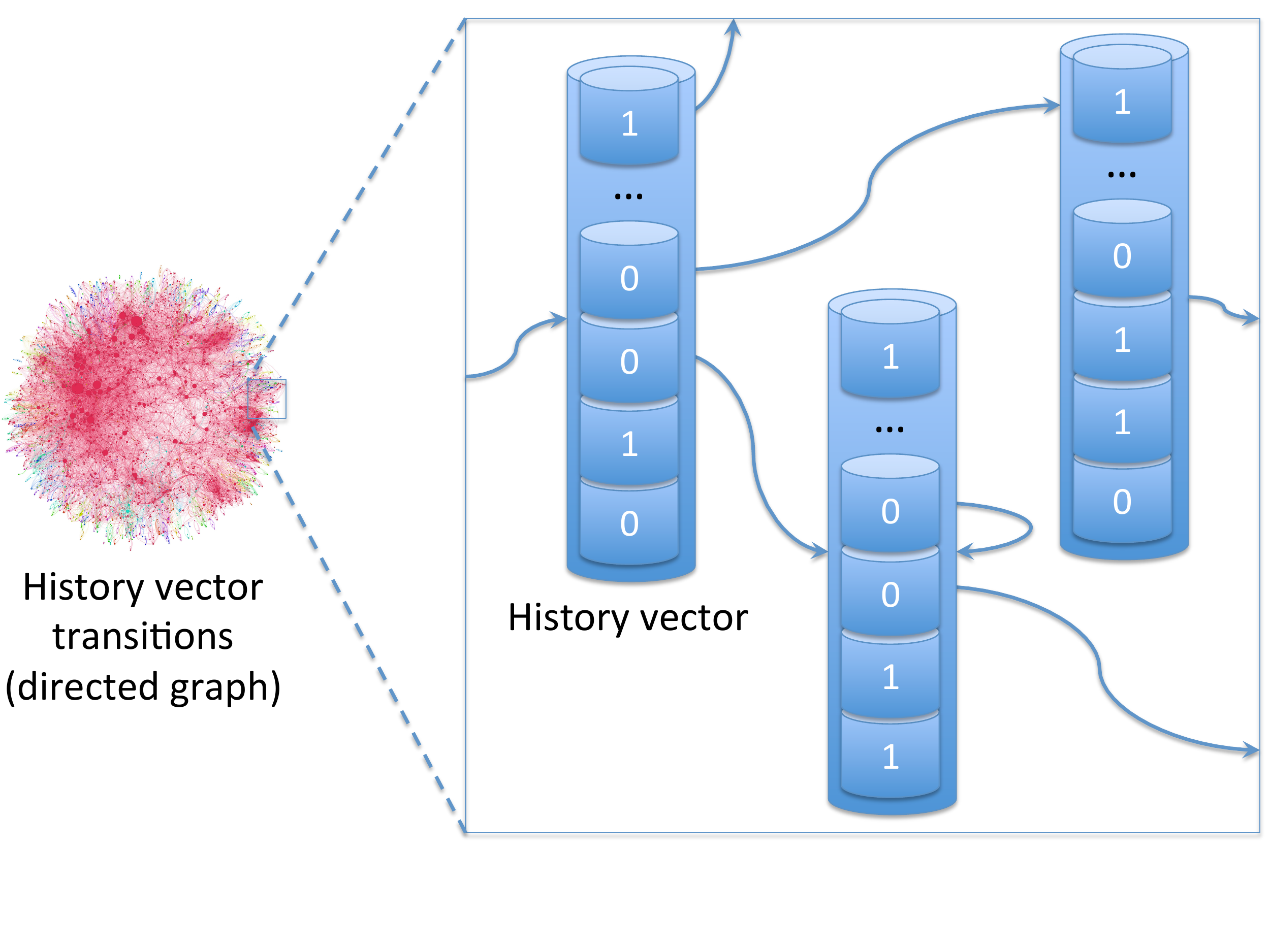}
  \vspace{-20pt}
  \caption{A conceptual summary of the adopted model which superimposes a Markovian model over a space of history vectors used to represent the medical state of a patient.}
  \label{f:concept}
  \vspace{-0pt}
\end{figure}

Given a patient's hospital admission history $H$ which comprises a sequence of admissions $a_i$:
\begin{align}
  H = a_1 \rightarrow a_2 \rightarrow \ldots \rightarrow a_n,
\end{align}
where each $a_i$ is a discrete variable whose value is a code from the International Statistical Classification of Diseases and Related Health Problems (ICD)~\cite{WHO2004}, the adopted model predicts the most likely future admission $a^*_{n+1}$ as:
\begin{align}
  a^*_{n+1} = \arg \max_{a \in A_\text{ICD}} p(H \rightarrow a),
  \label{e:prediction}
\end{align}
where $A_\text{ICD}$ is the set of all possible ICD codes. To make the estimation of the probability $p(H \rightarrow a)$ tractable, a patient's medical history $H$ is represented using a fixed length binary vector $v(H)$. This representation bears some resemblance to the bag of words representation frequently used in text analysis~\cite{BeykPhunAranVenk2015,BeykAranPhunVenk+2015} and which has since been successfully adapted to various other application domains too~\cite{Aran2012d,Aran2012f,RieuAran2015,SiviZiss2003}. Each element in $v(H)$ encodes the presence (value 1) or lack thereof (value 0) of a specific salient admission (i.e.\ ICD code) in $H$, save for the last element which captures jointly all non-salient admission types. As in~\cite{Aran2015g} saliency is determined by the frequency of the corresponding admission in the entire data corpus (n.b.\ different saliency criteria can be readily used instead, see Section~\ref{s:conc}). The probability $p(H \rightarrow a)$  in \eqref{e:prediction} is then estimated by superimposing a Markovian model~\cite{SukkKatzZhanRaun+2012,JackSharThomDuff+2003} on the space of history vectors which leads to $H \rightarrow a$ being interpreted as a transition from the state represented by $v(H)$ to the state represented by $v(H \rightarrow a)$. As usual the probabilities parameterizing the Markov model are learnt from a training data corpus. A conceptual illustration of the method is shown in Figure~\ref{f:concept}.

The key idea behind the described model is that it is the \emph{presence} of past complications which most strongly predicts future ailments~\cite{MudgKaspClaiRedf+2011,FrieJianElix2008,DharHsieLinBuen+2013,ButlKalo2012}, which allows for the space of states over which learning is performed to be reduced dramatically; in particular, this is achieved by employing a fixed length state representation and through binarization of its elements.

\subsection{Data and experimental protocol}
To make our results and conclusions directly comparable to those reported in the original paper which introduced the adopted history vector based method, we used the same large corpus of electronic medical records collected by a local hospital. The data set contains entries corresponding to nearly 3,000,000 hospital admissions of 40,000 patients.

Our primary goal here is to examine how the predictive performance of the history vector based model is affected by the choice of the number of salient admission types (see Section~\ref{ss:model}). As in~\cite{Aran2015g} we too assess the quality of a specific prediction by considering the rank of the ground truth admission type in the probability ordered list of predictions. Formally, let $a_t$ be the ground truth admission type which follows a particular history $H$. Then the rank $r$ of the ground truth admission type $a_t$ is given by number of admission types which the model predicts as following $H$ with at least the probability $p(H \rightarrow a_t)$:
\begin{align}
  r = | \left\{ a : a \in A_\text{ICD} \wedge p(H \rightarrow a) \geq p(H \rightarrow a_t)  \right\}  | .
\end{align}
We used the same granularity of codes the original work described in~\cite{Aran2015c,Aran2015g}.

\begin{figure*}[!t]
 \vspace{-0pt}
  \centering
  \footnotesize
   \includegraphics[width=1\textwidth]{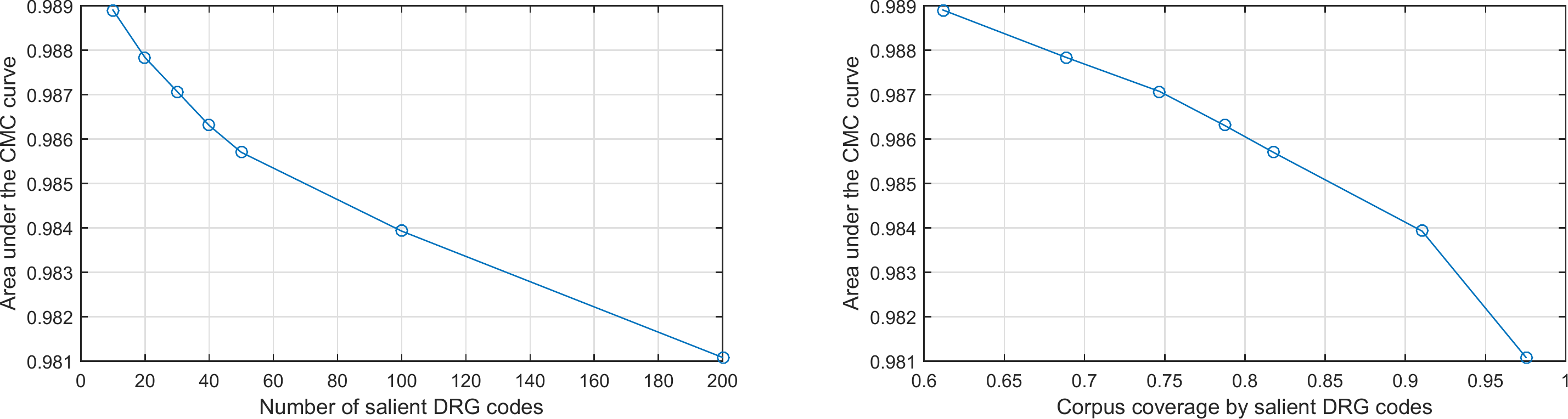}\\
   \begin{tabular}{p{0.23\textwidth}p{0.47\textwidth}p{0.3\textwidth}}
     & (a) & (b)\\
   \end{tabular}
   \vspace{-0pt}
   \caption{The normalized area under the cumulative match characteristic (CMC) curve. }
   \label{f:resCMC}
\end{figure*}

\begin{figure*}[t]
  \centering
  \footnotesize
   \includegraphics[width=1\textwidth]{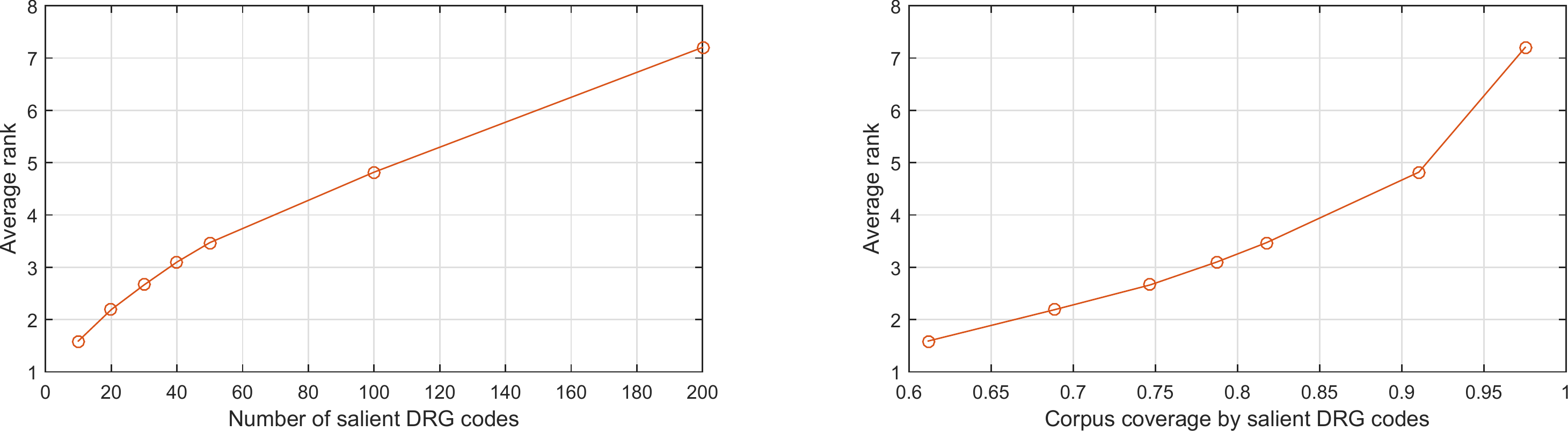}\\
   \begin{tabular}{p{0.23\textwidth}p{0.47\textwidth}p{0.3\textwidth}}
     & (a) & (b)\\
   \end{tabular}
   \vspace{-0pt}
   \caption{The average prediction rank of the correct admission type. }
   \label{f:resRank}
   \vspace{-0pt}
\end{figure*}

Furthermore, we adopt the usual `leave one out' evaluation protocol whereby the performance of the method is tested with each patient's data in turn and the model trained using the data of all other patients. To quantify the aggregate performance of the model for specific model parameter values (i.e.\ the number of salient admission types included in the history vector representation) we use two well known measures. These are the average rank (a special case of the average normalized rank~\cite{SaltMcGi1983} when the set of target matches is exactly equal to 1) and the normalized area under the cumulative match characteristic (CMC) curve. For each possible rank $r$ ($r=1\ldots n$, where $n$ is the worst possible rank, equal to the number of admission types), the CMC takes on the value equal to the proportion of predictions which predict the correct admission type at worst with the rank $r$~\cite{BollConnPankRath+2005}. The ideal performance results in the CMC having the value 1 across all ranks i.e.\ in each individual case the correct admission is ranked 1. The area under the curve is normalized so that it is equal to 1 in this ideal case.

\begin{figure*}[t]
  \centering
  \footnotesize
   \includegraphics[width=0.99\textwidth]{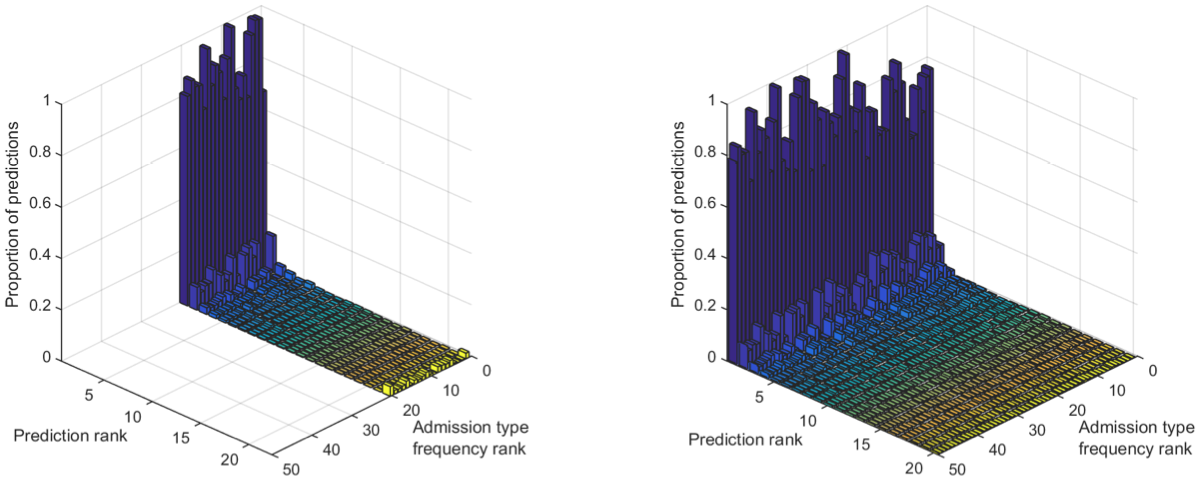}\\
   \begin{tabular}{p{0.23\textwidth}p{0.47\textwidth}p{0.3\textwidth}}
     & (a) & (b)\\
   \end{tabular}
   \caption{     Prediction rank histograms across different DRG codes using (a) 20 vs.\ (b) 50 salient admission types.   }
   \label{f:res3D}
   \vspace{-0pt}
\end{figure*}

\section{Results and Discussion}
We started by looking at the effect that changing the number of salient admission types, i.e.\ DRG codes with the corresponding (1-to-1) elements in the history vector, has on the area under the CMC curve. Our experimental results are captured by the plot in Figure~\ref{f:resCMC}(a). The plot can be readily seen to support the hypothesis put forward in Section~\ref{s:intro} that predicted a decay in the adopted model's prediction performance for an increasing number of explicitly modelled admission types. Notwithstanding this unwelcome qualitative observation, the major result is of a quantitative nature -- the rate of the aforementioned decay is very slow indeed. Like many other natural phenomena the decay exhibits a power-law form with the associated exponent value which differs from 1 only 5 parts in 100,000 i.e.\ it is equal to $1-0.5\times10^{-5}$. The practical significance of this finding is better appreciated by considering the plot in Figure~\ref{f:resCMC}(b). This plot shows the variation in the area under the CMC curve as a function of the coverage of the entire admissions data corpus by the salient DRG codes.  The outstanding performance of the adopted method is illustrated well by noting, for example, that the dimensionality of history vectors can be increased to explicitly model the number of most frequent DRG codes which cover over 91\% of the data , with the predictive performance of the method dropping by a mere 0.5\% as compared to the coverage of only 61\%. Even 98\% of data coverage results in a change of only 0.8\%. Recall that in the original paper the authors used 30 DRG codes which accounted for 75\% of the admissions in the corpus. Our results demonstrate that this was an overly conservative value.

We next examined the average prediction rank of the correct admission type,which offers further insight into the performance of the adopted method. As expected from the previous set of findings, the results summarized by the plots in Figs~\ref{f:resRank}(a) and~\ref{f:resRank}(b) corroborate the observation that an increase in the dimensionality of history vectors, a key parameter of the method, worsens performance. In this experiment this worsening is exhibited as an increase in the average rank (i.e.\ a greater number of incorrect predictions are made with a higher probability than the actual ground truth admission type). It is interesting to note the significance of what appears to be a much more rapid performance deterioration in terms of this performance measure in comparison with the area under the CMC curve discussed previously. For example, while the use of 200 vs.\ 10 most frequent DRG codes effects a reduction of only 0.5\% in the area under the CMC curve, the corresponding change in the average rank of the correct admission type increases fivefold (from approximately 1.5 for 10 salient DRG codes, to approximately 7.3 for 200 salient codes). The explanation for this apparent discrepancy is in fact reassuring as it demonstrates that the most dramatic changes in the predicted rank happen for predictions which are already not very good i.e.\ the small number of bad predictions become even worse, rather than good predictions becoming bad.

Lastly, to examine in additional detail how an increase in the number of explicitly modelled admission type affects predictions, we looked at prediction rank histograms for different DRG codes and the corresponding changes as their number was changed. Figure~\ref{f:res3D}(a) and~\ref{f:res3D}(b) contrast the histograms for 20 and 50 salient admission types. It is remarkable to observe that in both cases the histograms are virtually identical across different codes within the same model. Rather than being effected by sub-par histograms of the added DRG codes, the (small, as demonstrated previously) deterioration in predictive performance as the number of salient admission types is increased, is effected by slightly worse predictive performance uniformly distributed across different DRG codes. This is highly preferable in practice as it implies that for a fixed model complexity predictive power remains the same regardless of the patient's ailment. Were it otherwise, the predictions would be more difficult to interpret and the model complexity more challenging to set appropriately as the model's predictive performance would exhibit dependence on the nature of the health problems affecting a specific patient. 

\section{Summary and Conclusions}\label{s:conc}
In this paper we considered a recently proposed computational model which uses electronic medical records to predict future hospital admissions of a patient based on the patient's previous medical history. In particular we scrutinized the original authors' expectation that their preliminary results would scale as the number of explicitly modelled hospital admission types included in their history vector model is increased. Our experiments conducted on a real-world data corpus of nearly 3,000,000 admissions supports the authors' claims and demonstrate that the high prediction accuracy is maintained well even when the number of admission types explicitly included in the model is increased to account for 98\% of all admissions in the corpus.

Considering our findings, there are several well motivated directions for extending the adopted model. Firstly, the use of more granular DRG codes (i.e.\ codes corresponding to deeper hierarchical levels of the ICD's disease classification tree) should be investigated. Secondly, the practical significance of the predictions would greatly benefit from the use of temporal information. Lastly, alternative criteria for the choice of salient admission types should be examined; for example, instead of selecting these based on their frequency in the data, the criterion could be based on the number of different individuals affected (thereby eliminating the skew effected by conditions which may be experienced by a small number of people but which are chronic in nature), their association with mortality, or their cost.
\balance

\bibliographystyle{plain}
\bibliography{../../../../my_bibliography,../../../../oa_physiology} 

\end{document}